\documentclass[12pt]{article}    
\usepackage[dvips]{graphicx}

\begin{document}

\date{\today}
\title{Finite Temperature Nuclear Response in Extended RPA\footnote{
This work is supported in part by the U.S. DOE Grant No. DE-FG05-89ER40530.}}  
\author{Denis Lacroix$^{a)}$, Philippe Chomaz$^{a)}$ and Sakir Ayik$^{b)}$ \\
{\small $^{a)}$ {\it G.A.N.I.L., B.P. 5027, F-14021 Caen Cedex, France}} \\
{\small $^{b)}$ {\it Tennessee Technological University,
Cookeville TN38505, USA}}}
\maketitle

\begin{abstract}

{The nuclear collective response at finite temperature is investigated 
for the first time in the quantum framework of
the small amplitude limit of the extended TDHF approach,
including a non-Markovian collision term.
It is shown that the collision width satisfies 
a secular equation. By employing a Skyrme force,
the isoscalar monopole, isovector dipole and isoscalar quadrupole 
excitations in $^{40}Ca$ are calculated and 
important quantum features are pointed out.
The collisional damping due to decay into incoherent 2p-2h states is small 
at low temperatures but increases rapidly at higher temperatures. }
\end{abstract}

\vspace{3cm}


\section{Introduction}

After discovery of giant dipole resonance in 1947, 
much work has been done to understand the properties of nuclear
collective vibrations built on the ground state and excited states.
Most of these theoretical
investigations are based on the RPA theory which is
quite successful in describing the mean resonance energies and 
fragmentation of the excitation strengths at zero and finite temperatures.
However, the RPA approach, which is in fact the small amplitude limit of 
the TDHF theory,
is not suitable for describing damping 
of the collective excitations \cite{RingSchuck}. Damping arises mostly 
by mixing of the collective state with the nearby
complex states \cite{BerBro}.
The importance of the intrinsic compound nucleus
life-time has also been stressed 
\cite{moi,Ormand}.
As a result of the mixing with complex states, 
the excitation strength spread around
the mean resonance energy, and furthermore the damping width increases with 
the intrinsic temperature of the
system as observed in giant dipole resonances in $^{120} Sn$ \cite{Gar92}
and $^{208} Pb$
nuclei \cite{Ramak1,Ramak2,Hofmann}. In order to describe the nuclear 
collective response including
damping, it is necessary to go beyond the RPA theory by incorporating
coupling between the collective states and the doorway configurations.
There are essentially two different approaches for this purpose:(i)
Coherent mechanism due to coupling with low-lying surface modes which
provides an important mechanism for damping of giant resonances in particular
at low temperatures \cite{BerBorBro,Ormand}, 
(ii) Damping due to the coupling with incoherent
2p-2h states which is usually referred to as the collisional damping
\cite{DiToro}.
The small amplitude limit of the extended TDHF is an appropriate basis
for investigating collective response, in which damping due the incoherent
2p-2h decay is included in the form of a non-Markovian collision term
\cite{Ando,AyiDwo,Reinhard}.
Based on this approach, the incoherent contribution to damping
at finite temperature has been calculated in Thomas-Fermi approximation
in refs. \cite{BelAyi,Ayi,AyiYil}. 
Calculations using the Markovian 
limit of this semi-classical treatment, the so-called 
Boltzmann-Uehling-Uhlenbeck (BUU) approach,
are discussed by many authors (for a review see \cite{francesca}).
However, as far as the collective behavior of nuclei at moderate temperature
are concerned one may worry about the adequacy of semi-classical calculations
which neglect most of the quantum features but the Pauli principle.

In this work, we present a first quantal investigation
of  the nuclear collective response at zero and finite temperatures on the 
extended TDHF framework in small amplitude limit, 
which may be referred as an extended RPA approach.          
In this approach, in contrast to the semi-classical treatments, 
the shell effects are incorporated into the strength
distributions as well as the collisional damping widths. 
We point out that the damping widths should be calculated by solving
a secular equation. We compute the isoscalar monopole, isocalar quadrupole and
isovector dipole strength distributions in $^{40}Ca$ at finite temperatures
by employing an effective Skyrme force.

\section{ Collective response at finite temperature}

In the extended TDHF theory, the evolution of the single particle
density matrix $\rho(t)$ is determined by a transport equation
\cite{AyiYil,Abe,refETDHF1,Ayik,refETDHF2,refETDHF3,refETDHF4,refETDHF5},
\begin{equation}
i\hbar \frac{\partial}{\partial t}\rho-[h(\rho), \rho]= -\frac{i}{\hbar}
\int_{0}^{t} d\tau Tr_{2}
[v, G(t,t-\tau) F_{12}(t-\tau) G^{\dagger}(t,t-\tau)]
\end{equation}
where $h(\rho)$ is the mean-field Hamiltonian and the right hand side 
represents a non-Markovian collision term with
\begin{equation}
F_{12}=(1-\rho_1)(1-\rho_2) v \widetilde{\rho_1 \rho_2}-
\widetilde{\rho_1 \rho_2} v (1-\rho_1)(1-\rho_2)
\end{equation}
and $G(t,t-\tau)=T\cdot \exp [-\frac{i}{\hbar}
\int_{t-\tau}^{t} dt^{\prime} h(t^{\prime})]$ denotes the mean-field
propagator.
The small amplitude limit of the extended TDHF theory provides a 
suitable framework to describe 
collective vibrations including damping due to the coupling with 
incoherent 2p-2h excitations. The small deviations of the single-particle 
density matrix
$\delta \rho(t)=\rho(t)- \rho_{0}$ around a finite temperature equilibrium 
state $\rho_{0}$ is determined by
\begin{equation}
i\hbar \frac {\partial}{\partial t} \delta \rho - [h_{0}, \delta \rho] -
[\delta U+F, \rho_{0}] = I_{0}\cdot \delta \rho
\end{equation}
where $\delta U= (\partial U/\partial \rho)_{0}\cdot \delta \rho$ represents 
small deviations in the effective mean-field potential and 
$F({\bf r},t)=F({\bf r}) \exp(-i\omega t)+ h.c.$ is a one-body 
excitation
operator with harmonic time dependence. An explicit expression of 
the linearized form $I_0\cdot \delta \rho$ of the non-Markovian collision term
can be found in a recent publication \cite{AyiYil}. 

The linear response of the system to the external perturbation $F$ is 
determined by
expanding the small deviation $\delta \rho(t)$ in terms of finite temperature 
RPA modes,
\begin{equation}
\delta \rho(t)= \sum_{\lambda>0} \{ z_{\lambda}(t) \rho_{\lambda}^{\dagger}  + 
z_{\lambda}^{*}(t) \rho_{\lambda}\}
\end{equation}
where the finite temperature RPA modes, $\rho_{\lambda}^{\dagger}$ and
$\rho_{\lambda}$, are defined by
\begin{equation}
\hbar \omega_{\lambda} \rho_{\lambda}^{\dagger}- 
[h_{0}, \rho_{\lambda}^{\dagger }] -
[h_{\lambda}^{\dagger}, \rho_{0}]= 0.
\end{equation}
Here $\omega_{\lambda}$ is the mean-frequency of the RPA mode and
$h_{\lambda}^{\dagger}=
(\partial U/ \partial \rho)_{0}\cdot \rho_{\lambda}^{\dagger}$
represents the positive frequency part of the vibrating mean-field. 
It is convenient to
introduce 
the collective operators, $O_{\lambda}^{\dagger}$ and 
$O_{\lambda}$, associated with the RPA modes as
$\rho_{\lambda}^{\dagger }=[O_{\lambda}^{\dagger}, \rho_{0}]$ and
$\rho_{\lambda}=-[O_{\lambda}, \rho_{0}]$, and they are orthonormalized 
according to 
$Tr [O_{\lambda}, O_{\mu}^{\dagger}] \rho_{0}=\delta_{\lambda \mu} $. 
Substituting the 
expansion
(4)  and projecting by $O_{\lambda}$, we find that the amplitudes of the 
RPA modes execute forced harmonic motion,
\begin{equation}
-i\hbar \frac{d}{dt} z_{\lambda}+
(\hbar \omega_{\lambda}- \frac{i}{2}\Gamma_{\lambda}) z_{\lambda}
= <[O_{\lambda}, F]>_{0}
\end{equation}
where $<[O_{\lambda}, F]>_{0}=Tr[O_{\lambda}, F] \rho_{0}$ and
$\Gamma_{\lambda}$ denotes the collisional
damping width of the mode.
Here, we neglect a small shift of the mean frequency $\omega_{\lambda}$
arising from the principle value part of the collision term.
Solving this equation by Fourier transform, the response of the system to the 
external 
perturbation $F$ can be expressed as
\begin{equation}
\delta \rho (\omega)= R(\omega, T)\cdot F
\end{equation}
where $R(\omega, T)$ denotes the finite temperature extended
RPA response function 
including damping
\begin{equation}
R_{ij,kl}(\omega, T)= \sum_{\lambda>0}
\left( - \frac{<i| \rho_{\lambda}^{\dagger}|j><k|\rho_{\lambda}|l>}
{\hbar \omega-\hbar \omega_{\lambda}+\frac{i}{2}\Gamma_{\lambda}} +
 \frac{<k| \rho_{\lambda}^{\dagger}|l><i|\rho_{\lambda}|j>}
{\hbar \omega+\hbar\omega_{\lambda}+\frac{i}{2}\Gamma_{\lambda}} \right).
\end{equation}
The strength distribution of the RPA response is obtained by the imaginary part
of the response function,
\begin{eqnarray}
S(\omega, T)& =& -\frac{1}{\pi}Tr \{ F^{\dagger} Im R(\omega, T) F \} \\
&=& \frac{1}{\pi} \sum_{\lambda>0} 
\left \{ 
|<[O_{\lambda}, F]>_{0}|^2 D(\omega-\omega_{\lambda})-
|<[O_{\lambda}^{\dagger}, F]>_{0}|^2 D(\omega+\omega_{\lambda})
\right\}
\nonumber
\end{eqnarray}
where the sum goes over the positive frequency modes and
\begin{equation}
D(\omega-\omega_{\lambda})=
\frac{\Gamma_{\lambda}/2}
{(\hbar \omega-\hbar \omega_{\lambda})^2+(\Gamma_{\lambda}/2)^2}.
\end{equation}
The main features of the
strength function are usually discussed in terms of sum rules,
which are calculated from the RPA response as,
\begin{equation}
m_{k}(T) = \int_{0}^{\infty} \omega^{k}d \omega S(\omega, T) 
\end{equation}
However, due to the Lorentzian shape of the poles these moments are not
well defined for $k>1$. For a Hermitian excitation operator
the energy weighted sum rule for $k=1$ is not 
effected by the damping and it is given by,
\begin{equation}
m_{1}(T)= \sum_{\lambda >0} \omega_{\lambda}  |<[O_{\lambda}, F]>_{0}|^2. 
\end{equation}
For a multipole operator, the energy weighted sum rule leads to
$m_{1}(T)=\frac{1}{2} <[F^{\dagger}, [H, F]]>_{0}$ which
is exactly satisfied by the finite temperature RPA sum rule, 
as shown by Vautherin and Vinh Mau \cite{Vautherin}.  

In the Hartree-Fock basis the
finite temperature RPA equation reads,
\begin{eqnarray}
& (\hbar \omega_{\lambda}-\epsilon_{i}+\epsilon_{j})& <i| O_{\lambda}^{\dagger}|j>+ \\
& & \sum_{l \ne k} <ik|v|jl>_{A}
(n_{l}-n_{k}) <l| O_{\lambda}^{\dagger}|k>= 0   \nonumber
\end{eqnarray}
where $v=(\partial U/\partial \rho)_{0}$, the indices i,j,.. represent 
all single
particle quantum numbers including spin-isospin, and 
$n_{k}=1/[1+\exp(\epsilon_k-\epsilon_F)/T] $ denotes the
finite temperature Fermi-Dirac occupation numbers of the Hartree-Fock
states. At zero temperature these occupation numbers are zero or one,
so that the RPA operators
$O_{\lambda}^{\dagger}$, $O_{\lambda}$ have only
particle-hole and hole-particle matrix elements. At finite temperatures
the RPA functions involve more configurations including particle-particle
and hole-hole states. By associating a single index with the pair of
indices (i,j), the RPA functions can be regarded as a vector, and in this
manner eq.(13) can be expressed as an eigenvalue equation for finite 
temperature RPA modes \cite{Vautherin}.  
According to the small amplitude limit of the extended TDHF equation, 
the damping width of RPA modes due to decay into incoherent 2p-2h 
doorway excitations is given by \cite{AyiYil},
\begin{equation}
\Gamma_{\lambda}= \frac {1}{2}\sum
|<ij|[O_{\lambda},v]|kl>_{A}|^2 D_{ij,kl} 
[n_k n_l \bar{n}_i \bar{n}_j-
n_i n_j \bar{n}_k \bar{n}_l]
\end{equation}
where $\bar{n}_i=1-n_i$. In reference \cite{AyiYil}, 
neglecting the damping of the collective 
amplitude in the collision term, the energy conserving factor is taken as a 
sharp delta function as $D_{ij,kl}=
Im(\hbar \omega_{\lambda}-\epsilon_i-\epsilon_j+
\epsilon_k+\epsilon_l-i\eta)^{-1}$
Here, we take into account for depletion of the collective amplitude in the 
collision term by substituting 
$\omega_{\lambda}-\frac{i}{2} \Gamma_{\lambda}$ in place of
$\omega_{\lambda}$. Then, the factor takes a more appropriate Lorentzian form,
\begin{equation}
D_{ij,kl}= \frac{\Gamma_{\lambda}/2}
{(\hbar \omega_{\lambda}-\epsilon_i-\epsilon_j+\epsilon_k+\epsilon_l)^2
+(\Gamma_{\lambda}/2)^2} 
\end{equation}
Then, expression (14) becomes a secular equation for the damping width.
As we will see this self-consistency is of major importance in order to 
properly compute the collision width.
The collective mode damps out by mixing with the intrinsic states of increasing
complexity. The sequence of the complexity of the states can be
classified according to the exciton number as mixing with 
2p-2h, 3p-3h,$\cdots$,
Np-Nh,$\cdots$ states. The expression (14) contains only the mixing with 2p-2h
doorway states in accordance with the extended TDHF theory. In order to
incorporate the damping due to mixing with more complex states, the extended
TDHF should be improve by including higher order correlations beyond binary
collision term. The effect of the 
higher order mixing may be approximately
taken into account by introducing an appropriate decay width, 
$\Gamma_{ij,kl}$, of 2p-2h states
in the expression (15) 
\begin{equation}
D_{ij,kl}= \frac{(\Gamma_{\lambda}+\Gamma_{ij,kl})/2}
{(\hbar \omega_{\lambda}-\epsilon_i-\epsilon_j+\epsilon_k+\epsilon_l)^2
+((\Gamma_{\lambda}+\Gamma_{ij,kl})/2)^2} 
\end{equation}
Then, the secular equation can also be solved 
considering these higher order effects. 
 
\section{Results}

We calculate the isoscalar monopole, isoscalar quadrupole and isovector
dipole excitations in $^{40} Ca$ at several temperatures. We use the 
Skyrme interaction SGII for the Hartree-Fock and RPA calculations\cite{Van81}
and we neglect the temperature dependence of single particle energies and
wave functions. We determine the hole states by solving the Hartree-Fock
problem in coordinate representation. Then, the particle states are
generated by diagonalizing the Hartree-Fock Hamiltonian in a large harmonic
oscillator representation by including 12 major shells. In this manner, 
unbound continuum  states are approximately included in the RPA calculations.  
The RPA strength distributions  of
the monopole $F_{0}({\bf }r)=r^2$, 
dipole $F_{1}({\bf r})=\tau_{z}r Y_{10}(\hat{ {\bf r}})$ (in isospin symmetric
systems $N=Z$), and
quadrupole $F_{2}({\bf r})=r^{2} Y_{20}(\hat{{\bf r}})$ excitation operators
at temperatures $T=0,2,4$ MeV
are shown in figure 1. 
As seen from the top panel of figure 1, the monopole
strength at $T=0$ MeV exhibit a large Landau spreading over a broad 
energy region
$E=16-28$ MeV with an average energy $E=21.5$ MeV. 
The recent experimental
data also show a broad resonance around a peak value of
$17.5$ MeV\cite{Youngblood}. For increasing temperature, 
transition strength spread a broader
range towards lower energies. 
As shown in the middle panel, the strength 
distribution of isovector dipole
shows a weaker temperature dependence than monopole. At $T=0$, the dipole
strength is concentrated at range $E=16-23$ MeV. The Landau width
is large and is
spreading for increasing temperature. However, 
the average energy of the main peak
remains nearly
constant around $E=16.5$ MeV. The experimental data shows a broad resonance
at around $20$ MeV \cite{Berman} with a width close to 6 MeV.
As illustrated at the bottom panel of figure 1, the RPA result at $T=0$ MeV 
gives a very collective quadrupole mode peaked at $E=17.5$ MeV, 
which agrees well the experimental
finding of an average energy $17$ MeV \cite{Quadrupole} 
and the calculations of Sagawa and 
Bertsch \cite{Sagawa}. At higher temperatures in addition to p-h excitations, 
p-p and h-h excitations become possible. The p-p and h-h configurations
mainly change the strength distribution at low energy side at $E=4$ MeV. 
As a result, the giant resonance has less transition strength. 

Figure 2 illustrates the average energy $<E>=m_1/m_0$ (solid
lines),
and the ratios $\left( m_1/m_{-1} \right)^{1/2}$ (short dashed lines) and 
$\left( m_3/m_1 \right)^{1/2}$ (long dashed lines) of the monopole, 
dipole and quadrupole
excitations as a function of temperature. The behavior of the
average energy of the monopole
resonance is particularly interesting, since it may be related to the
compressibility coefficient of nuclear matter\cite{Bla80,Tre81}.
 
We obtain the collisional damping widths of the collective states by
calculating the expression (14) and solving the associated secular
equation by graphical method. We note that the sums over 
single particle states needed to evaluate the expresion (14), 
have been performed 
explicitely using the projection of the total spin, m, as one of 
the explicit quantum numbers as done in ref. \cite{phonon}. 
Figure 3 illustrates examples of the graphical solution for giant
dipole and quadrupole excitations at two temperatures $T=0,2$ MeV. 
In this figure,
the curves with dashed lines are obtained by calculating the right hand side
of (14) as a function of $\Gamma_{\lambda}=\Gamma_{in}$. The intersection of 
this curve with the diagonal line determines the solution. 
The effect of the damping width $\Gamma_{ij,kl}$ of 2p-2h states may be
approximately incorporated by taking a larger 
value of $\Gamma_{in}$ as indicated in eq.(16). 
As seen from figure 3, in most cases, the selfconsistent value of 
the damping width saturates very rapidly, and hence it 
is not modified very much by increasing $\Gamma_{in}$.
Figure 4 shows the damping 
widths as a function of temperature, that are averaged over several 
nearby states with strengths more than $10\%$ of the EWSR.
The results for monopole, dipole and quadrupole are indicated in the top,
middle and bottom panels, respectively. Collisional damping widths
are generally small at low temperatures, but rapidly grow for increasing
temperature. 
This increase appears to be more complex than the semi-classical quadratic
prediction. 
Depending upon the mode, the increase may be linear or may saturate. 

In order to understand this behavior, it is convenient
to write the expression (14) of the damping width as sum over energy bins in
energy
$E=\epsilon_i+\epsilon_j-\epsilon_k-\epsilon_l$ of 2p-2h states,
\begin{equation}
\Gamma_{\lambda}=\frac{1}{2} \sum_{E} 
g_{2p-2h}(E)\overline
{W}_{\lambda}(E) D(\hbar \omega_{\lambda}-E)
\end{equation}
where each bin has a small energy
interval $\Delta E$ around $E$, and $D(\hbar\omega_{\lambda}-E)$ is the
Lorentzian factor given by eq.(15).
Here $\overline{W}_{\lambda}(E)$ denotes the average
transition rate, 
\begin{equation}
\overline{W}_{\lambda}(E)=\frac{1}{g_{2p-2h}(E)} \sum_{\Delta E} 
|<ij|[O_{\lambda},v]|kl>_{A}|^2 
[n_k n_l \bar{n}_i \bar{n}_j-
n_i n_j \bar{n}_k \bar{n}_l].
\end{equation}
In this expression many terms are vanishing either due to
the selection rules or due to the Pauli blocking factors. 
The quantity $g_{2p-2h}(E)$ is the 
total number of 2p-2h states in the energy interval including
only those states which are not Pauli blocked 
and which have non-vanishing matrix elements of the transition rate, i.e., 
which can be coupled to the phonon quantum numbers.  
Figures 5 and 6 show the density of 2p-2h states and the average transition
rates as a function of the 2p-2h energy for dipole and quadrupole excitations
at $T=0$ MeV and $T=3$ MeV. In dipole mode, there is no odd parity 2p-2h
states available in the vicinity of the collective energy, hence the average
transition rate $\overline{W}_{\lambda}(E)$ and the density of states
vanish at zero temperature. As a result, 
the collisional damping of giant dipole in $^{40}Ca$ is zero at $T=0$ MeV.
This behavior is a particular quantum feature due to shell 
effects in the extended RPA calculation of double magic light nuclei,
and it can not be described in the framework of semi-classical approaches.  
In medium weight and heavy nuclei, in the vicinity of the GDR strength,
there are few odd parity 2p-2h configurations involving intruder states
associated with the spin-orbit coupling. As a result, we expect to find
a small finite damping of giant dipole resonance at zero temperature.
For increasing temperature, the available phase space becomes much 
larger and the collisional damping of the GMR and GDR increases. 
This is not the case for the GQR because the increase of the 
phase space is compensated by a reduction of the 
magnitude of average transition rates. As a result, the damping 
width of giant quadrupole appears to saturate above $T=3-4$ MeV.

Figure 7 shows the 
strength distributions including the collisional damping.
The giant dipole strength at $T=0$ MeV
is smoothed by performing an averaging with a Lorentzian weight with a
width of $0.5$ MeV. The excitation strengths become broader for increasing
temperature. The peak position of the monopole 
resonance does not change much, but the peak position of dipole slightly
shifts down and quadrupole slightly shifts up in energy. 
This is a signature of the reduction of the collectivity of those states
with temperature because the peak energy moves back towards the
single particle expectations. 

Figure 8 illustrates the FWHMs of the collective excitations 
at several temperatures. The widths of giant monopole
and giant dipole at $T=0$ MeV are mainly due to fragmentation of the collective
strength, i.e. the so-called Landau spreading,
which is about $4$ MeV in both cases. Since it is difficult to
extract well defined values, the FWHM of these modes at $T=0$ MeV 
are left open in the figure.
The total widths increase further by mixing of the collective mode with
incoherent 2p-2h excitations at higher temperatures. However, the total width
does not present the parabolic behavior predicted by semi-classical calculations.

In figure 9, long-dashed lines and solid lines show the integrated 
strengths over the energy interval $10-40$ MeV as a function of temperature
in the RPA and the extended RPA, respectively. As a reference, the total
strength $m_1$ is also indicated by short-dashed lines.
In the RPA calculations, the modes retain a high degree of collectivity
even at temperatures at $T=4-5$ MeV. However, in the extended RPA 
approach, as a result of damping, the excitation strength become 
broader and the collectivity diminishes for increasing temperature.    
 
\section{Conclusions}

We investigate isoscalar monopole, isoscalar quadrupole and isovector dipole 
excitations of $^{40}Ca$ at finite temperature in the basis of the small
amplitude limit of the extended TDHF. The extended TDHF goes beyond the
thermal RPA approach by including damping due to decay into incoherent
2p-2h excitations. We calculate 
the excitation strength distributions in a
self-consistent Hartree-Fock representation by employing a Skyrme force
with SGII parameters. At $T=0$, the monopole  and 
dipole strengths are fragmented and 
spread over a broad range because of the Landau damping, while quadrupole
exhibits a single peak structure. For increasing temperature, strength in all
cases becomes broader and hence the collectivity is reduced. The incoherent 
damping widths at low temperatures are, in general, small and thus leaving 
room for a possible coherence effect of doorway states in the description 
of the damping properties. At high temperature the collisional damping
becomes large and may even dominate the spreading width since the coherence 
effect is expected to diminish rather rapidly. For increasing temperature, 
the collisional damping predicted by the quantal calculations, evolves in
a more complex manner than the quadratic increase predicted by 
the semi-classical calculations. An interesting property of the collisional 
damping is that it may saturate for increasing temperature. In fact, our
calculations indicate that the damping width of giant quadrupole
saturates around $T=3-4$ MeV, however a saturation of the giant monopole 
and dipole modes is not visible at these temperatures.
There are important quantal effects in the collective behavior of a hot
nuclear system as illustrated in \cite{nous}.  
Investigations presented here, also indicates that, the quantal effects has a
large influence on the damping properties of collective 
excitations at low temperatures, which may even persist at relatively 
high excitations.
As illustrated in ref. \cite{AyiYil}, the magnitude of the collisional 
damping is rather sensitive
to the effective residual interactions, 
for which an
accurate information is not available. The effective Skyrme force is well
fitted to describe the nuclear mean-field properties, but not the in-medium
cross-sections and damping properties. Therefore, 
a systematic study of the effective interactions in this context is clearly 
called for.
However, our investigation, 
while remains semi-quantitative, gives a valuable insight on the 
quantal properties
of collective excitations at finite temperature. 
\vspace{2cm}

{\bf Acknowledgments}

One of us (S. A.) gratefully acknowledges GANIL Laboratory for a partial
support and warm hospitality extended to him during his visit to Caen.
This work is supported in part by the US DOE grant 
No. DE-FG05-89ER40530.

\begin{figure}[tbph]
\includegraphics*[height=16.6cm,width=15cm]{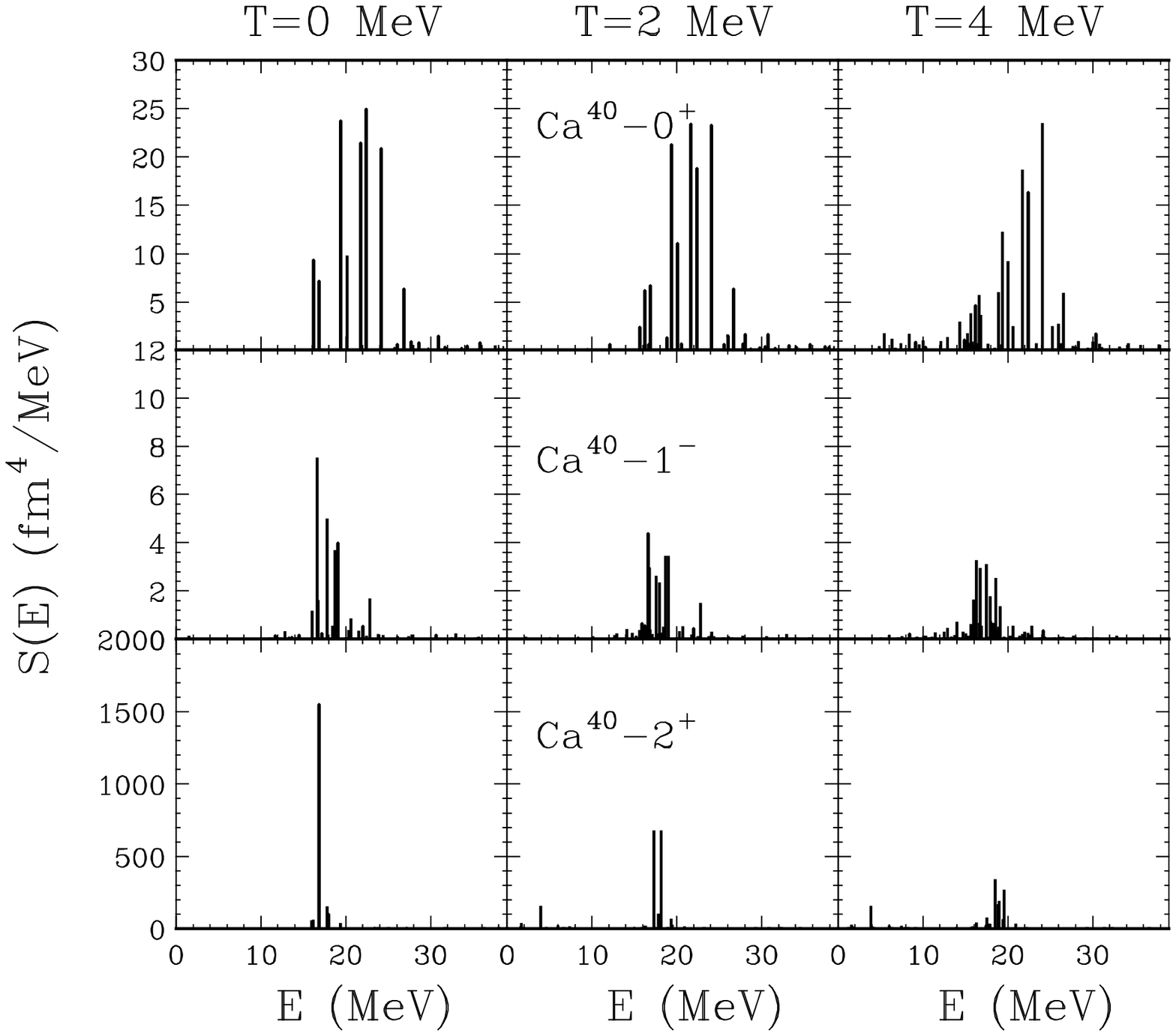}
\caption{RPA strength distributions in Ca$^{40}$ as a
function of the energy at temperatures $T = 0, 2, 4$ MeV
for isoscalar monopole $O^{+}$ (top), isovector dipole $1^{-}$
(middle) and isoscalar quadrupole $2^{+}$ (bottom) 
excitations.}
\label{fig:ca1}
\end{figure}
\begin{figure}[tbph]
\begin{center}
\includegraphics*[height=18cm,width=9cm]{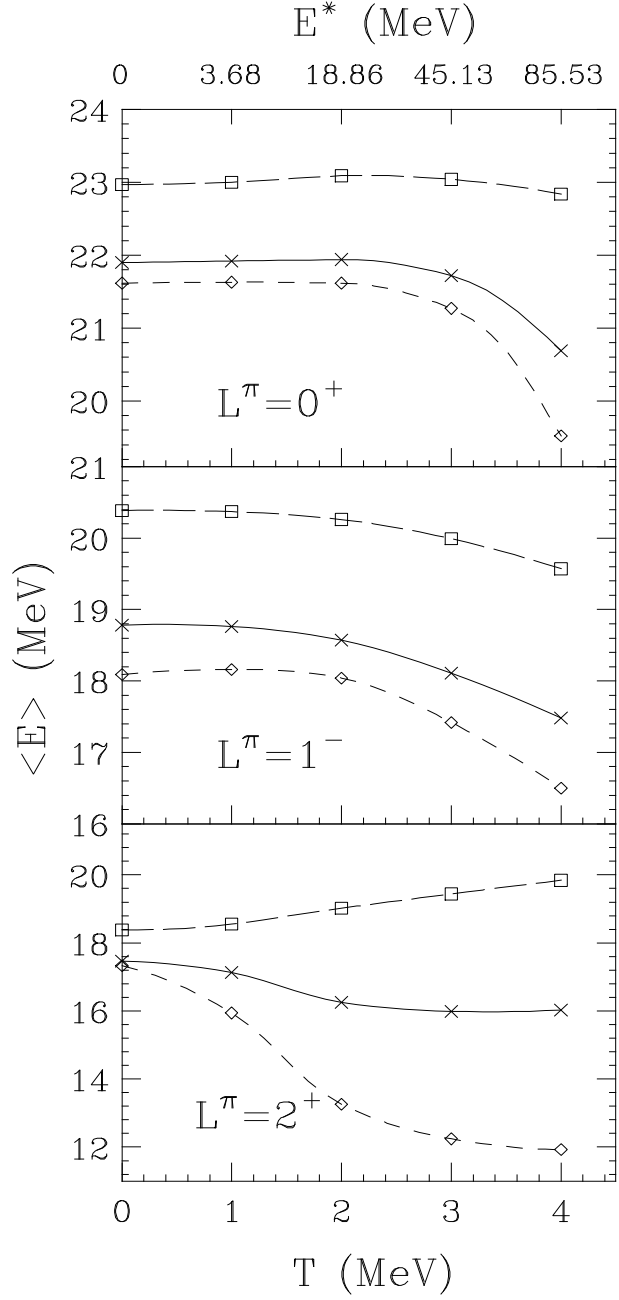}
\end{center}
\caption{Averaged energy of the monopole, dipole and quadrupole
excitations in $^{40}Ca$  $\displaystyle m_1/m_0$ (solid line)
and the moment ratios $\displaystyle \left( m_1/m_{-1} \right)^{1/2}$ 
(short dashed line) and $\displaystyle \left( m_3/m_{1} \right)^{1/2}$ 
(long dashed line) as a function of the temperature.}
\label{fig:hot4}
\end{figure}
\begin{figure}[tbph]
\begin{center}
\includegraphics*[height=11cm,width=14cm]{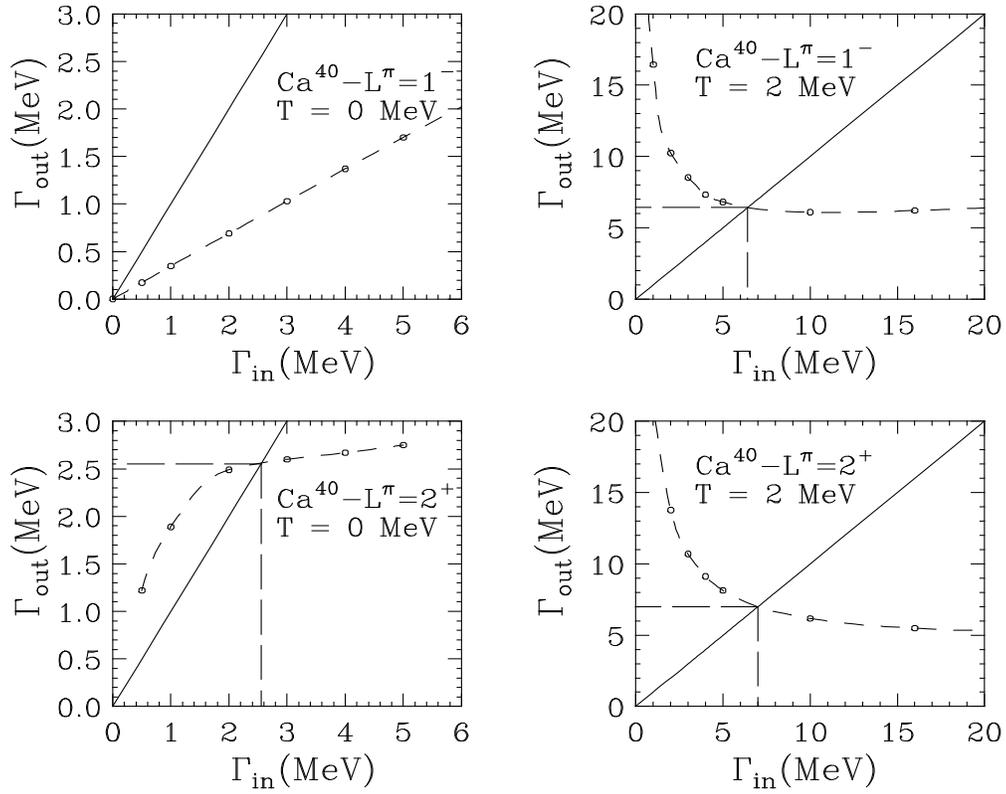}
\end{center}
\caption{Graphical solution of the secular equation for the damping width
for $L^\pi = 1^-$ (top) 
$L^\pi = 2^+$ (bottom) at T=0 MeV (left) and T=2 MeV (right). }
\label{fig:graphical}
\end{figure}
\begin{figure}[tbph]
\begin{center}
\includegraphics*[height=18.6cm,width=10cm]{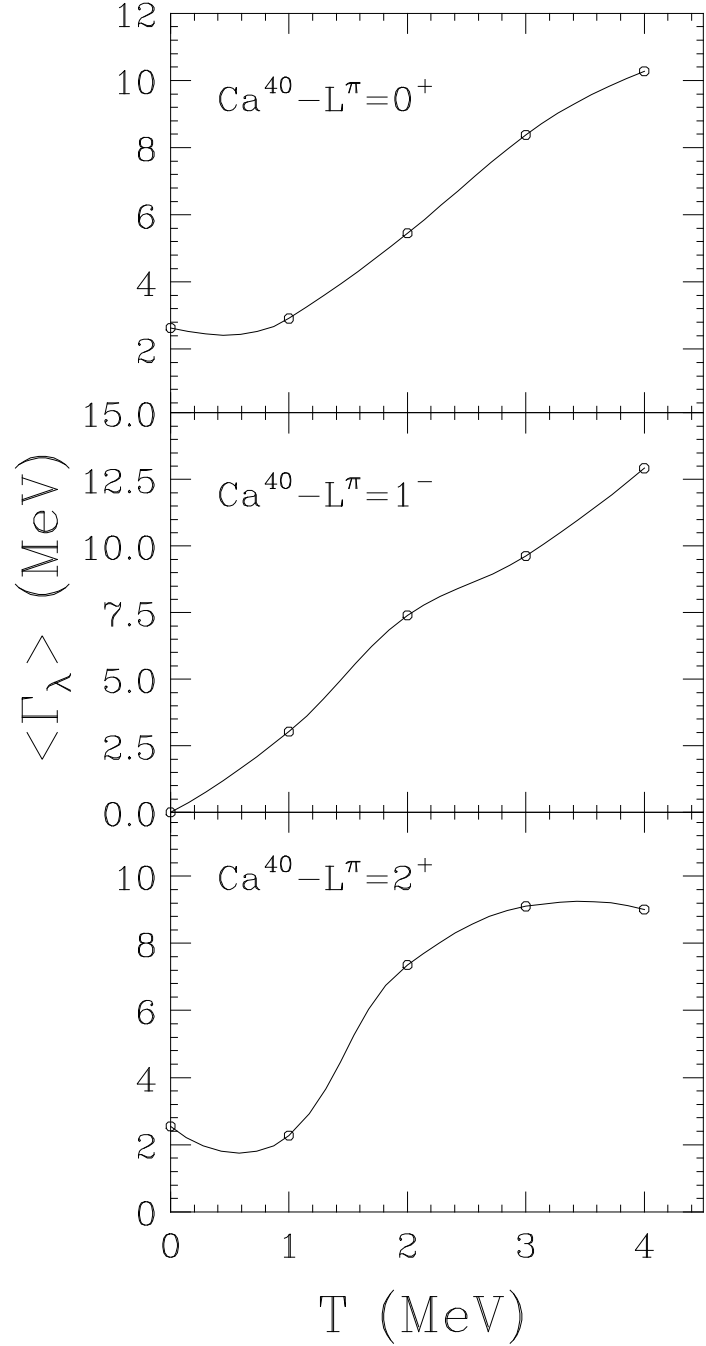}
\end{center}
\caption{Collisional damping widths that are averaged over nearby states 
with more than $10\%$ of the EWSR, for monopole, dipole and quadrupole modes
as a function of temperature.}
\label{fig:width_averaged}
\end{figure}

\begin{figure}[tbph]
\begin{center}
\includegraphics*[height=13cm,width=15cm]{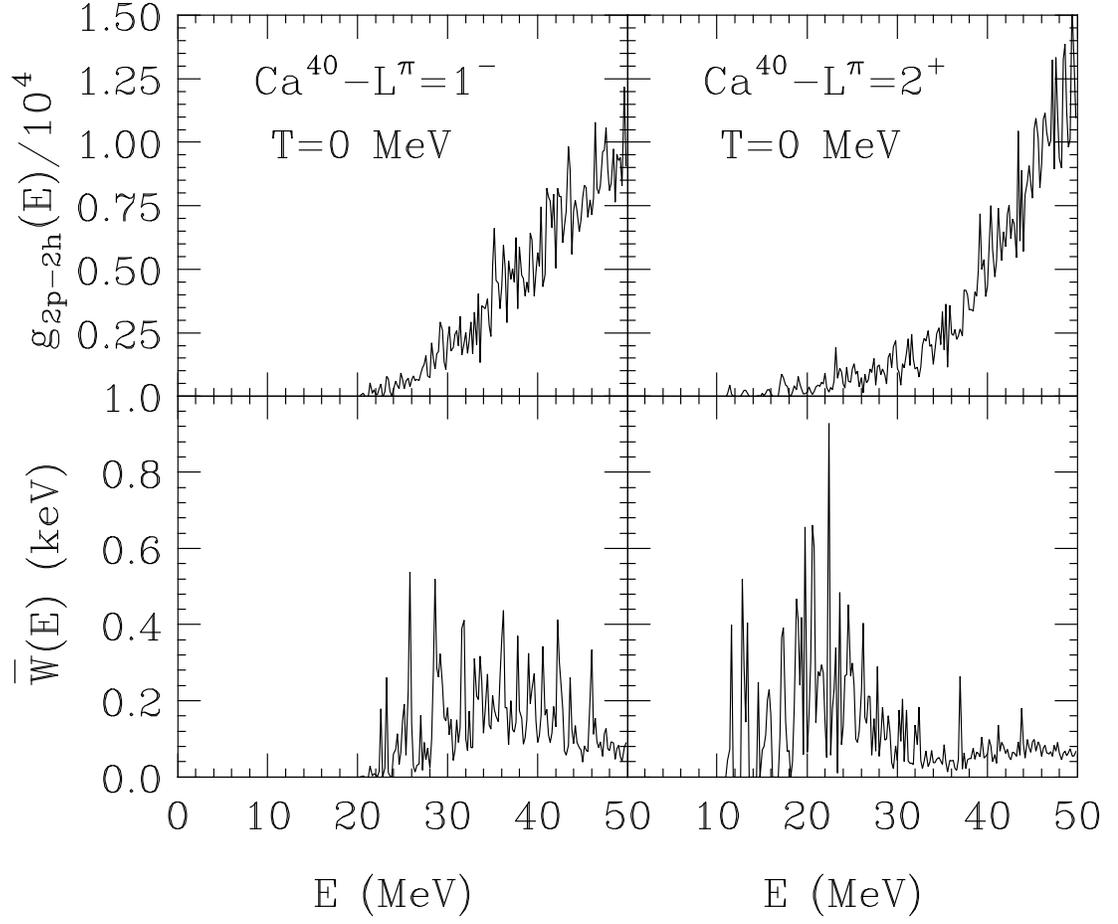}
\end{center}
\caption{Top: Energy dependence of the density of 2p-2h states $g_{2p-2h}$
for $L^\pi$=$1^-$(left) and $L^\pi$=$2^+$ (right) at zero temperature. 
Bottom: Averaged transition rate between collective states and 2p-2h states
as a function of the energy of $2p-2h$ states.}
\label{fig2:fine}
\end{figure}
\begin{figure}[tbph]
\begin{center}
\includegraphics*[height=13cm,width=14cm]{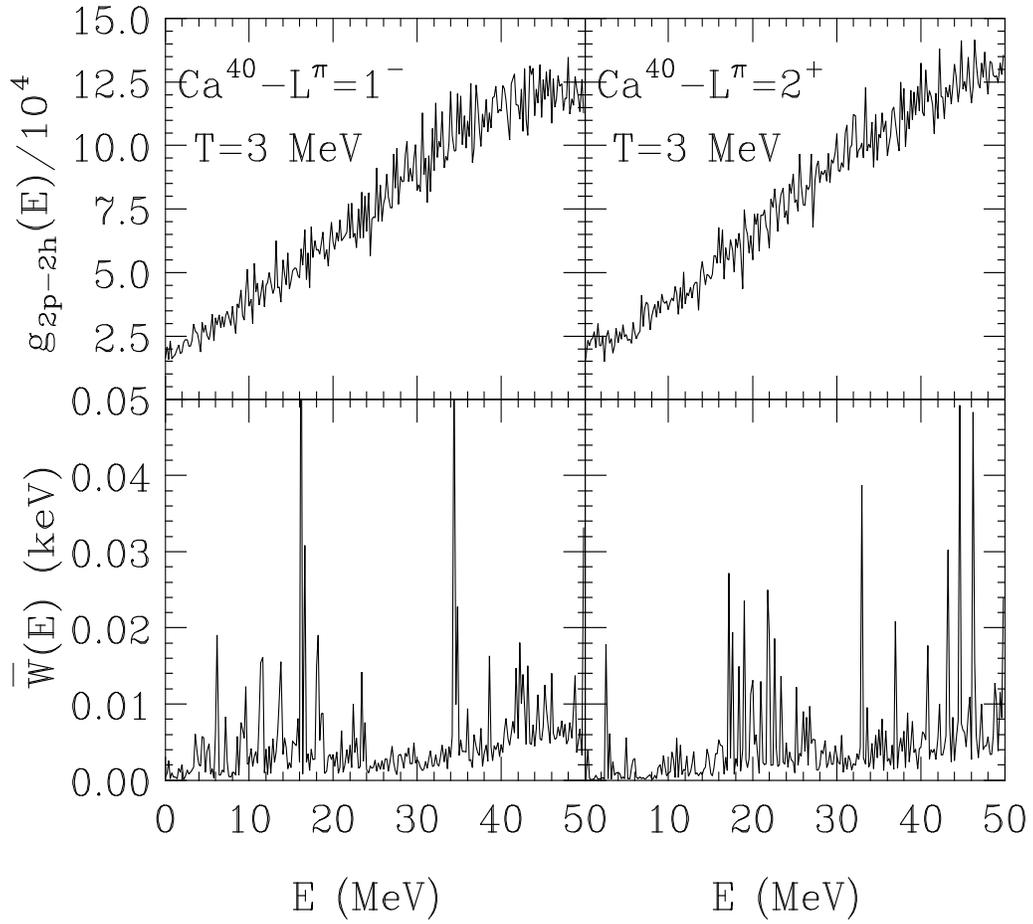}
\end{center}
\caption{The same as in figure 5, but at $T=3$ MeV.}
\label{fig3:fine}
\end{figure}

\begin{figure}[tbph]
\begin{center}
\includegraphics*[height=12.6cm,width=14cm]{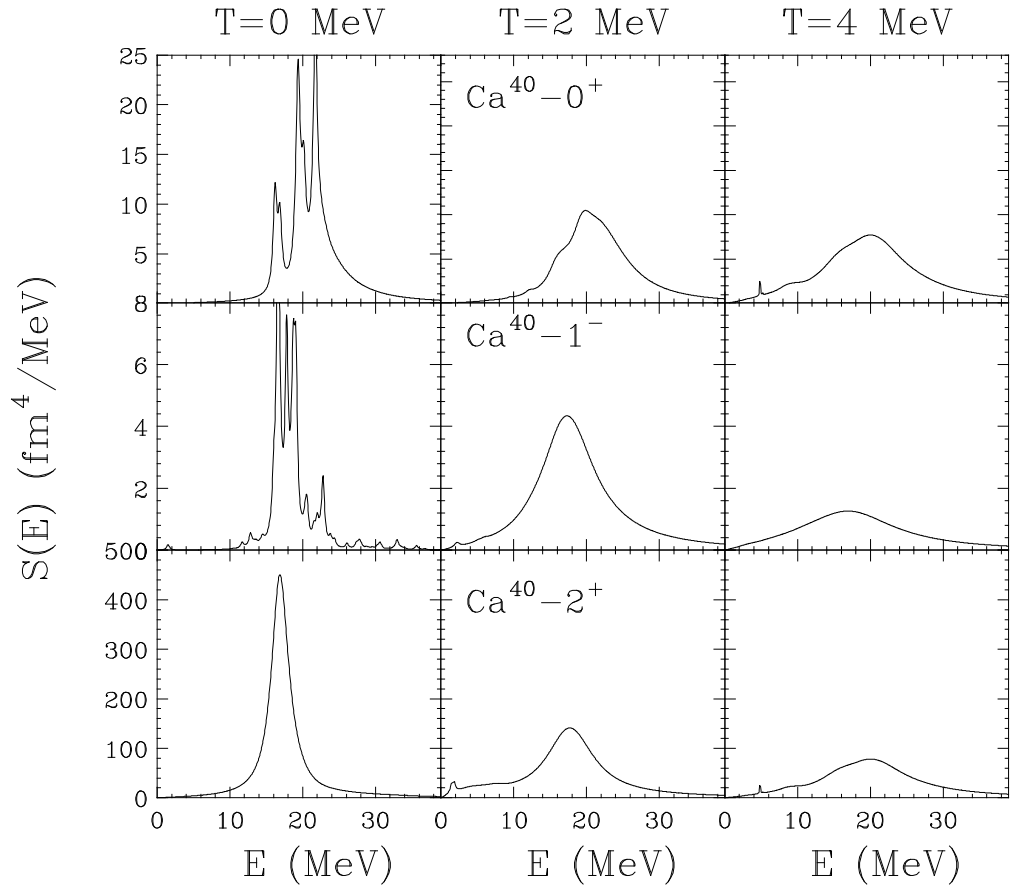}
\end{center}
\caption{The extended RPA strength distributions of the monopole, dipole and
quadrupole excitations as a function of temperature.}
\label{fig:streng_coll}
\end{figure}
\begin{figure}[tbph]
\begin{center}
\includegraphics*[height=18cm,width=10cm]{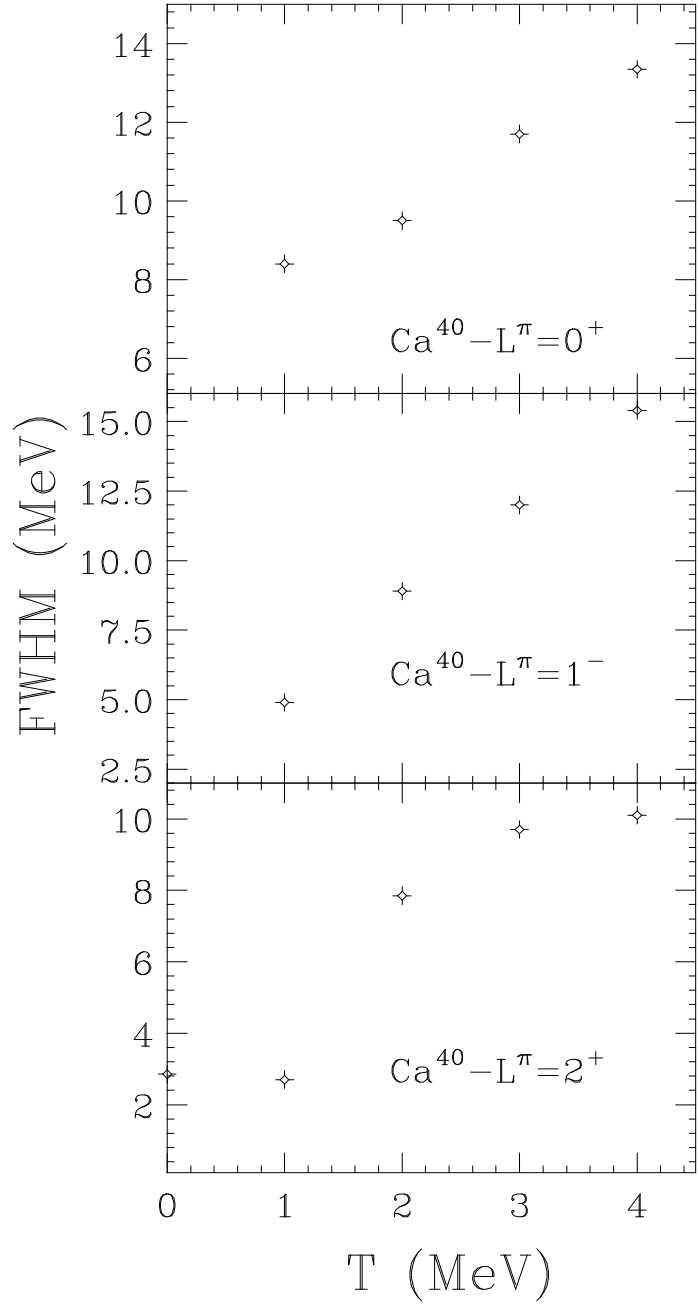}
\end{center}
\caption{The full widths at half maximum of the strength
distributions at several temperatures.} 
\label{fig:fwhm_coll}  
\end{figure}
\begin{figure}[tbph]
\begin{center}
\includegraphics*[height=18cm,width=10cm]{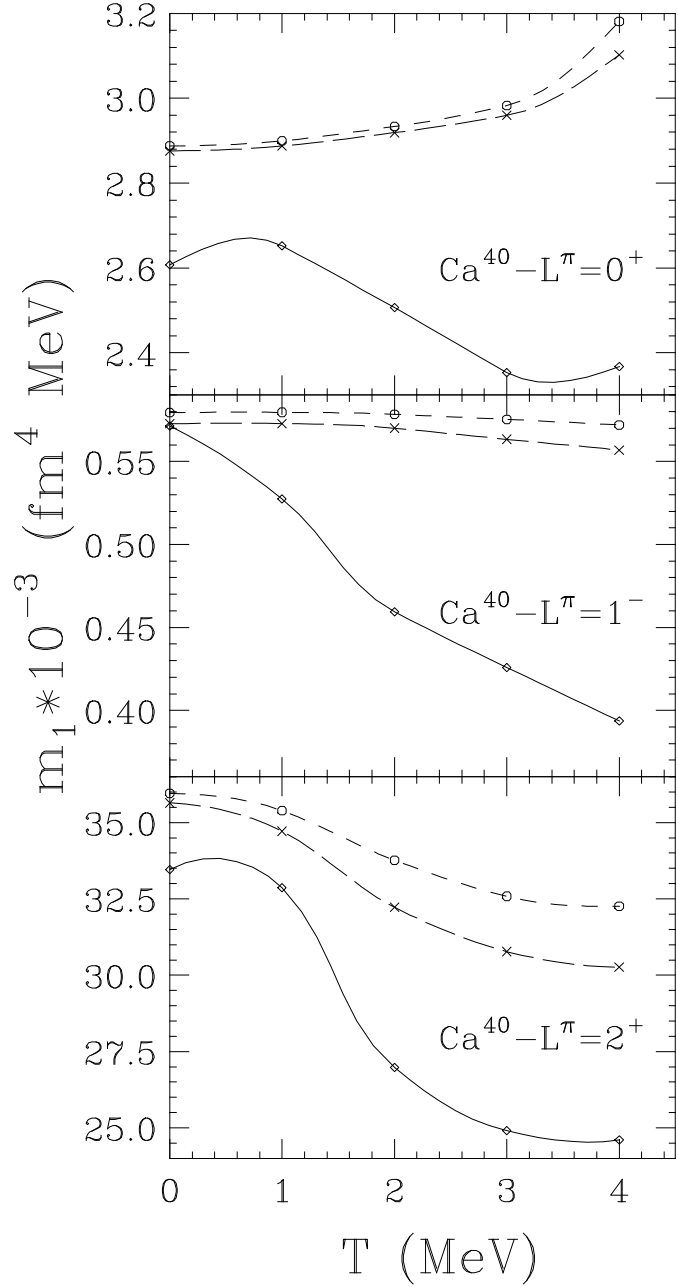}
\end{center}
\caption{Integrated strengths over the energy interval $10-40$ MeV in the RPA
(long-dashed line) and the extended RPA (solid line). The total strengths
$m_1$ are plotted as a reference by short dashed lines.} 
\label{fig:m1}  
\end{figure}

\end{document}